\documentclass[preprint,amsmath,amssymb,11pt,english,aps,showkeys,showpacs]{revtex4}
\usepackage{graphicx}
\usepackage{graphics}
\usepackage{amsmath}
\usepackage{dcolumn}
\usepackage{amssymb}
\usepackage{bm}

\begin{document}
\title{Noninertial effects on a Dirac neutral particle inducing an analogue of the Landau quantization in the cosmic string spacetime}
\author{Knut Bakke}
\email{kbakke@fisica.ufpb.br}
\affiliation{Departamento de F\'isica, Universidade Federal da Para\'iba, Caixa Postal 5008, 58051-970, Jo\~ao Pessoa, PB, Brazil.}

\begin{abstract}
We discuss the behaviour of external fields that interact with a Dirac neutral particle with a permanent electric dipole moment in order to achieve relativistic bound states solutions in a noninertial frame and in the presence of a topological defect spacetime. We show that the noninertial effects of the Fermi-Walker reference frame induce a radial magnetic field even in the absence of magnetic charges, which is influenced by the topology of the cosmic string spacetime. We then discuss the conditions that the induced fields must satisfy to yield the relativistic bound states corresponding to the Landau-He-McKellar-Wilkens quantization in the cosmic string spacetime. Finally we obtain the Dirac spinors for positive-energy solutions and the Gordon decomposition of the Dirac probability current.
\end{abstract}


\maketitle

\section{Introduction}

\label{sec:1}

In recent decades, the Landau quantization \cite{landau} has been investigated in studies of Bose-Einstein condensates \cite{l1}, the quantum Hall effect \cite{l2}, linear topological defects \cite{l3,l4}, and neutral particles \cite{er,lin}. The Landau quantization describes the interaction between a charged quantum particle and a uniform magnetic field; the resulting discrete spectrum of energies corresponds to the quantization of a charged particle in cyclotron orbits. The Landau quantization was also investigated in relativistic systems by Rabi \cite{l5}, Berestetskii {\it et al.} \cite{l6}, Jackiw \cite{l7} and Balatsky {\it et al.} \cite{l8}. The interest in the relativistic Landau quantization has motivated discussions on spin-nematic states \cite{l9}, the finite-temperature problem \cite{l10}, and the quantum Hall effect \cite{l11}. More recently, the relativistic Landau quantization has been extended to condensed-matter systems described by the Dirac equation \cite{l12}, and to neutral particles \cite{bf5}. Following these studies of relativistic quantum systems, the relativistic Landau quantization for neutral particles has also been investigated in a noninertial reference frame \cite{b3} and in the Lorentz symmetry-violation background \cite{bbs}.

The objective of this work is to obtain bound states solutions for the Dirac equation by discussing the behaviour of external fields that interact with a Dirac neutral particle with a permanent electric dipole moment in a particular noninertial frame, and in the presence of a topological defect spacetime. Studies of noninertial effects on quantum systems are well documented. The most widely known noninertial effects on quantum systems are the Sagnac effect \cite{sag,sag5} and the Mashhoon effect \cite{r3}, related to geometric phases, which arise in interferometric experiments. Other studies of noninertial effects and geometric phases have obtained the Berry phase \cite{r7}, the influence of gravitational effects on quantum interferometry \cite{r9}, and an analogue of the Aharonov-Casher effect \cite{bf4}. Related to bound state solutions, the Page-Werner \emph{et al} coupling \cite{r1,r2,r4} is an interesting quantum effect due to noninertial effects. Additional work on noninertial effects on quantum systems includes studies of scalar fields \cite{r8}, Dirac fields \cite{r10}, Lorentz transformations \cite{r5}, the weak-field approximation \cite{r6}, the confinement of a neutral particle to a two-dimensional quantum dot \cite{b4}, the Landau-He-McKellar-Wilkens quantization \cite{b}, and the Landau-Aharonov-Casher quantization \cite{bf16}. 

Here, without invoking magnetic charges, we show that the noninertial effects of the Fermi-Walker reference frame induce a radial magnetic field, which depends on the topology of the cosmic-string spacetime. In addition, we discuss the conditions that the induced fields must satisfy to yield relativistic bound states corresponding to the relativistic Landau-He-McKellar-Wilkens quantization \cite{bf5} in the cosmic string spacetime. Finally we obtain the Dirac spinors for positive-energy solutions and the Gordon decomposition of the Dirac probability current.

The structure of this paper is the following: in Section II, we discuss the behavior of the Dirac spinors and external fields in a noninertial frame in the cosmic string spacetime; in Section III, we find the conditions that must be satisfied by the induced fields to yield the relativistic bound states; in section IV, we obtain the Dirac spinors for positive-energy solutions and the Gordon decomposition of the Dirac probability current; in section V, we present our conclusions.

\section{spinors and external fields in a noninertial frame and in the cosmic string spacetime}
\label{sec:2}

We start out by introducing the cosmic string spacetime background, and building the noninertial reference frame in which we we will obtain a radial magnetic field even without magnetic charges. The cosmic string is a topological defect \cite{kat,def,def2,def3,staro} whose appearance is considered to be due to a symmetry breaking related to phase transitions in the early Universe \cite{def}. With the mathematical tools of general relativity, we can describe the cosmic
string spacetime by a line element of the following form ($\hbar=c=1$ in our units):
\begin{eqnarray}
ds^{2}=-d\bar{t}^{2}+d\bar{\rho}^{2}+\eta^{2}\bar{\rho}^{2}d\bar{\phi}^{2}+d\bar{z}^{2},
\label{1.1}
\end{eqnarray}
where the parameter $\eta$, related to the deficit angle, is defined as $\eta=1-4\varpi G/c^{2}$, with $\varpi$ being the linear mass density of the cosmic string. In the cosmic string spacetime, the parameter related to the deficit angle can only assume values in which $0<\eta<1$, and the azimuthal angle varies in the interval $0\,\leq\,\bar{\phi}\,<\,2\pi$. It is worth mentioning another context involving topological defects and the spatial part of the line element of the cosmic string spacetime. In the solid state physics, the spatial part of line element of the cosmic string, called disclination, can be obtained from the Katanaev-Volovich approach \cite{kat}, which uses geometric models equivalent to those of general relativity to describe such linear topological defects in crystalline solids as dislocations, disclinations, and dispirations \cite{moraesG2}. However, we need to take into account that the values of the parameter $\eta$ in crystalline solids can differ from that of the line element of the cosmic string spacetime. In crystalline solids, the parameter $\eta$ can assume both values $0<\eta<1$ and $\eta>1$. All values where $\eta>1$ corresponding to an anti-cone with negative curvature \cite{kat,def3,moraesG2}. Returning to the geometry described by the line element (\ref{1.1}), we have a non-null curvature given by the following curvature tensor:
\begin{eqnarray}
\label{curv}
R_{\rho,\varphi}^{\rho,\varphi}=\frac{\left(1-\eta\right)}{4\eta}\,\delta_{2}(\vec{r}),
\end{eqnarray}
where $\delta_{2}(\vec{r})$ is the two-dimensional delta function. The spacetime curvature given by Eq.~\eqref{curv} is concentrated on the symmetry axis of the cosmic string, all other spacetime positions having null curvature. This configuration defines a conical singularity \cite{staro}.

To study noninertial effects on a Dirac neutral particle interacting with external fields, let us make the simple coordinate transformation $\bar{t}=t;\,\,\,\,\bar{\rho}=\rho;\,\,\,\,\bar{\phi}=\varphi+\omega\,t;\,\,\,\,\bar{z}=z$, where the parameter $\omega$ corresponds to the constant angular velocity of the rotating frame. With this transformation, the line element (\ref{1.1}) of the cosmic string spacetime becomes
\begin{eqnarray}
ds^{2}=-\left(1-\omega^{2}\eta^{2}\rho^{2}\right)\,dt^{2}+2\omega\eta^{2}\rho^{2}d\varphi\,dt+d\rho^{2}+\eta^{2}\rho^{2}d\varphi^{2}+dz^{2}.
\label{1.5}
\end{eqnarray}
This line element is defined for radial coordinates in the range:
\begin{eqnarray}
0\,<\rho\,<\frac{1}{\eta\omega}.
\label{1.5a}
\end{eqnarray}

The line element in Eq.~(\ref{1.5}) is not well defined for $\rho\geq1/\omega\eta$, since radial coordinates satisfying this inequality would place the particle outside the light-cone, that is, would call for speeds greater than the velocity of light \cite{landau3}.  The range (\ref{1.5a}) constrains the wave function of the Dirac particle: the wave function $\psi\left(x\right)$ must vanish as $\rho\rightarrow1/\omega\eta$.

Henceforth, to obtain a relativistic analogue of Landau quantization, we will focus on the behaviour of the external fields that interact with a Dirac neutral particle with a permanent electric dipole moment. To this end, let us build a noninertial reference frame for the observers where we can define the Dirac spinors locally, as prescribed by the spinor theory in curved space \cite{weinberg}. We have seen that the line element (\ref{1.1}) and consequently the line element (\ref{1.5}) have a non-null curvature concentrated on the symmetry axis of the cosmic string. In a curved spacetime background, spinors must be defined locally in such a way that each spinor transforms according to infinitesimal Lorentz transformations $\psi'\left(x\right)=D\left(\Lambda\left(x\right)\right)\,\psi\left(x\right)$, where $D\left(\Lambda\left(x\right)\right)$ corresponds to the spinor representation of the infinitesimal Lorentz group, and $\Lambda\left(x\right)$ corresponds to the local Lorentz transformations \cite{weinberg}. A local reference frame can be built by means of a noncoordinate basis $\hat{\theta}^{a}=e^{a}_{\,\,\,\mu}\left(x\right)\,dx^{\mu}$ whose components  $e^{a}_{\,\,\,\mu}\left(x\right)$ are called tetrads and satisfy the relation: $g_{\mu\nu}\left(x\right)=e^{a}_{\,\,\,\mu}\left(x\right)\,e^{b}_{\,\,\,\nu}\left(x\right)\,\eta_{ab}$ \cite{weinberg,naka}. The Greek indices denote the coordinates of the cosmic string spacetime, while the latin indices denote the local reference frame of the observers. The tensor $\eta_{ab}=\mathrm{diag}(- + + +)$ is the Minkowski tensor. The tetrads $e^{a}_{\,\,\,\mu}\left(x\right)$ have an inverse defined as $dx^{\mu}=e^{\mu}_{\,\,\,a}\left(x\right)\,\hat{\theta}^{a}$, where the tetrads and the inverse of the tetrads are related via $e^{a}_{\,\,\,\mu}\left(x\right)\,e^{\mu}_{\,\,\,b}\left(x\right)=\delta^{a}_{\,\,\,b}$ and $e^{\mu}_{\,\,\,a}\left(x\right)\,e^{a}_{\,\,\,\nu}\left(x\right)=\delta^{\mu}_{\,\,\,\nu}$. From these definitions of the local reference frames, the external fields in a background having a non-null curvature and noninertial effects are defined as \cite{rt2}:
\begin{eqnarray}
F^{\mu\nu}\left(x\right)=e^{\mu}_{\,\,\,a}\left(x\right)\,e^{\nu}_{\,\,\,b}\left(x\right)\,F^{ab}\left(x\right),
\label{1.6}
\end{eqnarray}
where $F^{ab}\left(x\right)$ corresponds to the electromagnetic tensor defined in the rest frame of the observers. The components to the $F^{ab}\left(x\right)$ tensor are defined by: $F^{0i}=-E^{i}$ and $F^{ij}=-\epsilon^{ijk}\,B_{k}$.

Let us construct a local reference frame for the observers defined at each instant in the rest frame of the observers, with no rotation of the spatial axis: 
\begin{eqnarray}
\hat{\theta}^{0}=dt;\,\,\,\hat{\theta}^{1}=d\rho;\,\,\,\hat{\theta}^{2}=\eta\omega\rho\,dt+\eta\rho\,d\varphi;\,\,\,\hat{\theta}^{3}=dz,
\label{1.8}
\end{eqnarray}
the definition $\hat{\theta}^{0}=e^{0}_{\,\,\,t}\left(x\right)\,dt$ ensuring that the local reference frame~(\ref{1.8}) is the observer's rest frame at each instant. Note that there is no rotation of the spatial components of the noncoordinate basis $\hat{\theta}^{i}$  $\left(i=1,2,3\right)$, which corresponds to a Fermi-Walker reference frame \cite{misner}. The most important characteristic of the Fermi-Walker reference frame is that we can observe noninertial effects due to the action of external forces without any effects from arbitrary rotations of the local spatial axis of the reference frame of the observers. Our interest is that the noninertial effects in this nonrotating frame can induce external fields without torques or external forces on the electric dipole moment of the Dirac neutral particle. As example, consider a uniform electric $\vec{E}_{\mathrm{rf}}=\mathbf{E}_{0}\,\hat{z}$ in the rest frame of the observers, that is, $E^{3}=\mathbf{E}_{0}$. From (\ref{1.6}) and (\ref{1.8}), the field configuration in the noninertial frame and in the cosmic string background is \cite{b}
\begin{eqnarray}
E^{z}=\mathbf{E}_{0};\,\,\,\,\,\,\,B^{\rho}=-\omega\,\eta\,\mathbf{E}_{0}\,\rho.
\label{1.12}
\end{eqnarray}

The relativistic Landau-He-McKellar-Wilkens quantization was proposed in Ref. \cite{bf5} by applying the duality transformation on the relativistic Landau-Aharonov-Casher setup. In this way, the conditions that the field configuration must satisfy in order to achieve the relativistic Landau quantization in the He-McKellar-Wilkens setup \cite{hmw} are the electrostatic conditions, the absence of torque on the electric dipole moment, and the presence of a uniform effective magnetic field perpendicular to the plane on which the neutral particle moves. From the field configuration induced by the noninertial effects of the Fermi-Walker reference frame (\ref{1.12}), we can see that both electrostatic conditions and the absence of torque on the electric dipole moment are satisfied. Before defining the effective magnetic field, we note that to describe the quantum dynamics of a neutral particle with a permanent electric dipole moment it is convenient to start from the relativistic Dirac theory and add a nonminimal coupling into the Dirac equation (in Minkowski spacetime) given by $i\gamma^{\mu}\,\partial_{\mu}\rightarrow\,i\gamma^{\mu}\,\partial_{\mu}+i\frac{d}{2}\,\Sigma^{\mu\nu}\,\gamma^{5}\,F_{\mu\nu}\left(x\right)$ \cite{anan,sil}. From the introduction of this nonminimal coupling, Anandan \cite{anan} showed that there exists an effective gauge potential 
\begin{eqnarray}
A_{\mu}^{\mathrm{eff}}=\left(\vec{d}\cdot\vec{E},\,\vec{d}\times\vec{B}\right),
\label{1.13}
\end{eqnarray}
where the vector $\vec{d}=d\,\vec{\sigma}$ (here the $\vec{\sigma}=\left(\sigma^{1},\,\sigma^{2},\sigma^{3}\right)$ are the Pauli matrices) corresponds to the permanent electric dipole moment of the neutral particle. In this way, the effective magnetic field is defined in the form:
\begin{eqnarray}
\vec{B}_{\mathrm{eff}}=\vec{\nabla}\times\vec{A}_{\mathrm{eff}}=\vec{\nabla}\times\left[\hat{n}\times\vec{B}\right],
\label{1.14}
\end{eqnarray}
with $\hat{n}$ being a unit vector on the direction of the permanent electric dipole moment of the neutral particle. 

With the field configuration~(\ref{1.12}) we then find that $\vec{B}_{\mathrm{eff}}=-2\omega\,\eta\,\mathbf{E}_{0}$, that is, we have a uniform effective magnetic field perpendicular to the plane on which the neutral particle moves. Therefore, the conditions for relativistic Landau-He-McKellar-Wilkens quantization established in \cite{bf5} are satisfied by the field configuration induced by noninertial effects on the Fermi-Walker reference frame (\ref{1.12}). Note, by applying the duality transformation as suggested in \cite{bf5} in the relativistic regime and in \cite{lin} in the nonrelativistic regime, that the field configuration for achieving the Landau quantization in the He-McKellar-Wilkens setup \cite{hmw} is characterized by the presence of a radial magnetic field produced by a magnetic charge density. In this work, the field configuration (\ref{1.12}) is also characterized by the presence of a radial magnetic field, but without the hypothesis of the existence of magnetic charges. Moreover, we can see the influence of the topology of the defect on the field configuration given by the presence of the parameter $\eta$ in the expression of the radial magnetic field, which is due to the noninertial effects of the FermiWalker reference frame.

\section{Dirac equation and relativistic Landau-He-McKellar-Wilkens quantization in the cosmic string spacetime}
\label{sec:3}

We now discuss the interaction between the external fields and the electric dipole moment of the neutral particle, which leads to relativistic bound-state solutions for the Dirac equation in a noninertial frame in the background of the cosmic string spacetime. We want to find the relativistic analogue of Landau quantization satisfying the constraint~(\ref{1.5a}) on the radial coordinate. To this end, we go back to the discussion of spinor theory in curved space \cite{weinberg} in Section~\ref{sec:2} and recall that the spinors are defined in the observer's local reference frame. Therefore, to properly write the equation of motion for a Dirac particle in a curved spacetime, we must replace the partial derivative of the spinors with the covariant derivative \cite{bd,naka,schu}. The covariant derivative of a spinor is defined by the equality $\nabla_{\mu}=\partial_{\mu}+\Gamma_{\mu}\left(x\right)$, where $\Gamma_{\mu}=\frac{i}{4}\,\omega_{\mu ab}\left(x\right)\,\Sigma^{ab}$ corresponds to the spinorial connection~\cite{bd,naka}. The components of the spinorial connection can be obtained with the tetrads (\ref{1.8}) by solving the Cartan structure equations in the absence of the torsion field $d\hat{\theta}^{a}+\omega^{a}_{\,\,\,b}\wedge\hat{\theta}^{b}=0$, where the symbol $\wedge$ denotes the wedge product, while the operator $d$ is the exterior derivative \cite{naka}). The following Dirac equation therefore describes the relativistic quantum dynamics of the neutral particle with permanent electric dipole moment interacting with external magnetic and electric fields in the cosmic string spacetime:
\begin{eqnarray}
i\gamma^{\mu}\,\partial_{\mu}\psi+i\gamma^{\mu}\,\Gamma_{\mu}\left(x\right)\psi+i\frac{d}{2}\,\Sigma^{\mu\nu}\,\gamma^{5}\,F_{\mu\nu}\left(x\right)\psi=m\psi,
\label{2.1}
\end{eqnarray}
where $d$ is the permanent electric dipole moment of the neutral particle, $F_{\mu\nu}\left(x\right)$ is the electromagnetic field tensor, and $\Sigma^{ab}=\frac{i}{2}\left[\gamma^{a},\gamma^{b}\right]$ (the indices $a,b,c=0,1,2,3$ indicate the local reference frame). The $\gamma^{a}$ matrices corresponds to the standard Dirac matrices defined in the Minkowski spacetime \cite{greiner}:
\begin{eqnarray}
\gamma^{0}=\hat{\beta}=\left(
\begin{array}{cc}
1 & 0 \\
0 & -1 \\
\end{array}\right);\,\,\,\,\,\,
\gamma^{i}=\hat{\beta}\,\hat{\alpha}^{i}=\left(
\begin{array}{cc}
 0 & \sigma^{i} \\
-\sigma^{i} & 0 \\
\end{array}\right);\,\,\,\,\,\,\gamma^{5}=\left(
\begin{array}{cc}
0 & I \\
I & 0 \\	
\end{array}\right);\,\,\,\,\Sigma^{i}=\left(
\begin{array}{cc}
\sigma^{i} & 0 \\
0 & \sigma^{i} \\	
\end{array}\right),
\label{2.3}
\end{eqnarray}
where $I$ is the $2\times2$ identity matrix and $\vec{\Sigma}$, the spin vector. The matrices $\sigma^{i}$ are the Pauli matrices and satisfy the relation $\left(\sigma^{i}\,\sigma^{j}+\sigma^{j}\,\sigma^{i}\right)=2\,\eta^{ij}$. The $\gamma^{\mu}$ matrices given in the Dirac equation (\ref{2.1}) are related to the $\gamma^{a}$ matrices via $\gamma^{\mu}=e^{\mu}_{\,\,\,a}\left(x\right)\gamma^{a}$.

Next, we solve the Cartan structure equations in the absence of torsion to obtain four non-null components of the connection 1-form $\omega^{a}_{\,\,\,b}=\omega_{\mu\,\,\,\,b}^{\,\,\,a}\left(x\right)\,dx^{\mu}$:  $\omega_{t\,\,\,2}^{\,\,\,1}\left(x\right)=-\omega_{t\,\,\,1}^{\,\,\,2}\left(x\right)=-\omega\eta$ and $\omega_{\varphi\,\,\,2}^{\,\,\,1}\left(x\right)=-\omega_{\varphi\,\,\,1}^{\,\,\,2}\left(x\right)=-\eta$. From these four non-null components of the connection 1-form, we can calculate all components of the spinorial connection $\Gamma_{\mu}\left(x\right)$ and obtain $i\gamma^{\mu}\,\Gamma_{\mu}=i\frac{\gamma^{1}}{2\rho}$ \cite{b,b3}. The Dirac equation describing the interaction of the permanent electric dipole moment of the neutral particle with the induced fields (\ref{1.12}) in the cosmic string background is given by the following expression: 
\begin{eqnarray}
i\frac{\partial\psi}{\partial t}=m\hat{\beta}\psi+i\omega\frac{\partial\psi}{\partial\varphi}-i\hat{\alpha}^{1}\left(\frac{\partial}{\partial\rho}+\frac{1}{2\rho}\right)\psi-i\frac{\hat{\alpha}^{2}}{\eta\rho}\frac{\partial\psi}{\partial\varphi}-i\hat{\alpha}^{3}\frac{\partial\psi}{\partial z}+id\hat{\beta}\vec{\alpha}\cdot\vec{B}\psi+d\hat{\beta}\vec{\Sigma}\cdot\vec{E}\psi.
\label{2.6}
\end{eqnarray}

Since the operators $\hat{J}_{z}=-i\frac{\partial}{\partial\varphi}$ \cite{schu} and $\hat{p}_{z}=-i\frac{\partial}{\partial z}$ commute with the Hamiltonian given in the right-hand-side of the equation (\ref{2.6}), we write the solution of the Dirac equation (\ref{2.6}) in terms of the eigenvalues of the operators $\hat{p}_{z}=-i\frac{\partial}{\partial z}$ and $\hat{J}_{z}=-i\frac{\partial}{\partial\varphi}$ \footnote{Ref. \cite{schu} has shown that the $z$-component of the total angular momentum operator is given by $\hat{J}_{z}=-i\frac{\partial}{\partial\varphi}$ whose eigenvalues are $j=l+\frac{1}{2}$, where $l=0,\pm1,\pm2,\ldots$.}: 
\begin{eqnarray}
\psi=e^{-i\mathcal{E}t}\,e^{i\,j\,\varphi}\,e^{ikz}\,\left(
\begin{array}{c}
\xi\left(\rho\right)\\
\chi\left(\rho\right)\\	
\end{array}\right),
\label{2.7}
\end{eqnarray} 
where $j=l+\frac{1}{2}$, $l=0,\pm1\pm2,\ldots$, while $k$ is a constant, $\xi=\left(\xi_{+}\,\,\,\xi_{-}\right)^{T}$ and $\chi=\left(\chi_{+}\,\,\,\chi_{-}\right)^{T}$ are two-component spinors, where $\sigma^{3}\xi_{+}=\xi_{+}$, $\sigma^{3}\xi_{-}=-\xi_{-}$ and likewise for $\chi_{\pm}$. Substituting this solution into the Dirac equation (\ref{2.6}) we obtain two coupled equation of $\xi\left(\rho\right)$ and $\chi\left(\rho\right)$. The first coupled equation is
\begin{eqnarray}
\left[\mathcal{E}-m+\omega\left(l+\frac{1}{2}\right)-d\mathbf{E}_{0}\,\sigma^{3}\right]\xi=\left[-i\,\sigma^{1}\frac{\partial}{\partial\rho}-\frac{i\,\sigma^{1}}{2\rho}-id\omega\eta\mathbf{E}_{0}\,\rho\,\sigma^{1}+\frac{\sigma^{2}}{\eta\rho}\left(l+\frac{1}{2}\right)+k\,\sigma^{3}\right]\chi,
\label{2.8}
\end{eqnarray}
while the second one is
\begin{eqnarray}
\left[\mathcal{E}+m+\omega\left(l+\frac{1}{2}\right)+d\mathbf{E}_{0}\,\sigma^{3}\right]\chi=\left[-i\,\sigma^{1}\frac{\partial}{\partial\rho}-\frac{i\,\sigma^{1}}{2\rho}+id\omega\eta\mathbf{E}_{0}\,\rho\,\sigma^{1}+\frac{\sigma^{2}}{\eta\rho}\left(l+\frac{1}{2}\right)+k\,\sigma^{3}\right]\xi.
\label{2.9}
\end{eqnarray}

Eliminating $\chi$ from Eq. (\ref{2.9}) and considering the electric dipole moment parallel to the $z$ axis, we obtain two decoupled equations for $\xi_{+}$ and $\xi_{-}$, which we label $\xi_{s}$, where $s=\pm1$, so that $\sigma^{3}\xi_{s}=\pm\xi_{s}=s\xi_{s}$. We can then rewrite the two decoupled equations in a more compact form representing two radial equations:
\begin{eqnarray}
\frac{d^{2}\xi_{s}}{d\rho^{2}}+\frac{1}{\rho}\frac{d\xi_{s}}{d\rho}-\frac{\zeta^{2}_{s}}{\eta^{2}\rho^{2}}\,\xi_{s}-\delta^{2}\,\rho^{2}\,\xi_{s}+\beta_{s}\,\xi_{s}=0\qquad(s=\pm),
\label{2.10}
\end{eqnarray}
where $\zeta_{s}$ is an effective angular momentum, defined by the equality
\begin{eqnarray}
\zeta_{s}=l+\frac{1}{2}\left(1-s\right)+\frac{s}{2}\left(1-\eta\right).
\label{2.11}
\end{eqnarray}

The other two parameters on the right-hand side of Eq.~\eqref{2.10} are 
\begin{eqnarray}
\delta=d\,E_{0}\,\omega\,\eta;\,\,\,\,\,\,\,\,\beta_{s}=\left[\mathcal{E}+\omega\left(l+\frac{1}{2}\right)\right]^{2}-\left[m+s\,d\,E_{0}\right]^{2}-2s\,\delta\,\frac{\zeta_{s}}{\eta}-2\,\delta.
\label{2.12}
\end{eqnarray}

We have set $k=0$ in Eqs.~(\ref{2.10})~and (\ref{2.12}) because there is no torque on the dipole moment \cite{er}. Our problem is therefore reduced to the plane of motion of the Dirac neutral particle, the solutions of the radial equations~(\ref{2.10}) being of the form
\begin{eqnarray}
\xi_{s}\left(\rho\right)=e^{-\frac{\delta\,\rho^{2}}{2}}\,\left(\delta\,\rho^{2}\right)^{\frac{\left|\zeta_{s}\right|}{2\eta}}\,F_{s}\left(\rho\right).
\label{2.12a}
\end{eqnarray}

This result substituted on the right-hand side of Eq.~(\ref{2.10}), we obtain the following second order differential equation:
\begin{eqnarray}
\mu\,F_{s}''+\left[\frac{\left|\zeta_{s}\right|}{\eta}+1-\mu\right]\,F_{s}'+\left[\frac{\beta_{s}}{4\delta}-\frac{\left|\zeta_{s}\right|}{2\eta}-\frac{1}{2}\right]\,F_{s}=0,
\label{2.14}
\end{eqnarray}
where $\mu=\delta\,\rho^{2}$. 

Equation (\ref{2.14}) is the Kummer equation and $F_{s}\left(\mu\right)=F\left[\frac{\left|\zeta_{s}\right|}{2\eta}+\frac{1}{2}-\frac{\beta_{s}}{4\delta},\frac{\left|\zeta_{s}\right|}{\eta}+1,\mu\right]$ is the confluent hypergeometric function or Kummer function of first kind \cite{abra}. It has been widely discussed in the literature \cite{landau2,abra,b,f2,f3,lin} that the radial part of the wave function becomes finite everywhere when the parameter $\frac{\left|\zeta_{s}\right|}{2\eta}+\frac{1}{2}-\frac{\beta_{s}}{4\delta}$ is equal to a non-positive integer number, which turns the confluent hypergeometric series into an $n$th-degree polynomial. Taking into account that the line element (\ref{1.5}) is valid only for $0<\rho<\frac{1}{\omega\eta}$, we assume that 
\begin{eqnarray}
dE_{0}\ll\omega\eta.
\label{2.14a}
\end{eqnarray}

This restriction on the intensity of the electric field yields a probability amplitude that becomes very small for $\rho\geq1/\omega\eta$ because the parameter $\mu=\delta\rho^{2}\ll1$ for $\rho\rightarrow\frac{1}{\omega\eta}$. With no loss of generality, therefore, the radial wave function can be considered as being normalized inside the range $0<\rho<\frac{1}{\omega\eta}$. The relativistic energy levels for the bound state solutions of the Dirac equation for a neutral particle with permanent electric dipole moment are 
\begin{eqnarray}
\mathcal{E}_{n,\,l}=\sqrt{\left(m+s\,d\,E_{0}\right)^{2}+4dE_{0}\omega\eta\,\left(n+\frac{\left|\zeta_{s}\right|}{2\eta}+s\frac{\zeta_{s}}{2\eta}+1\right)}-\omega\left[l+\frac{1}{2}\right].
\label{2.15}
\end{eqnarray}

Equation~(\ref{2.15}) is the relativistic analogue of the Landau-He-McKellar-Wilkens quantization. We have therefore achieved the relativistic Landau quantization for the He-McKellar-Wilkens setup \cite{hmw} without assuming the hypothesis of the existence of a density of magnetic charges producing a radial magnetic field. Our derivation relies on the noninertial effects of the Fermi-Walker reference frame, which induce a radial magnetic field similar to the one in the Landau-He-McKellar-Wilkens quantization proposed in \cite{lin,bf5}. We also have that the presence of the topological defect breaks the degeneracy of the relativistic analogue of the Landau levels \cite{f2,f3,bf6} given in (\ref{2.15}). For $\eta\rightarrow1$, we recover the relativistic Landau-He-McKellar-Wilkens quantization induced by noninertial effects in the Minkowski spacetime obtained in \cite{b3}. For $\omega\rightarrow0$, the relativistic energy levels of bound states (\ref{2.15}) vanish because there are no noninertial effects inducing the field configuration (\ref{1.12}).

Hence, we have seen that the relativistic Landau-He-McKellar-Wilkens quantization induced by the noninertial effects of the Fermi-Walker reference frame in the cosmic string spacetime can be achieved if the intensity of the electric field satisfies the condition (\ref{2.14a}). For other values of the intensity of the electric field, the amplitude of probability cannot be sufficiently small for values of the radial coordinate $\rho\geq1/\omega\eta$, therefore we cannot consider the wave function being normalized inside the physical region of the spacetime $0\,<\,\rho\,<\,1/\omega\eta$.

Consider next the nonrelativistic limit of the energy levels (\ref{2.15}). For $m>>s\,d\,E_{0}$ and $m>>4dE_{0}\omega\eta\left(n+\frac{\left|\gamma_{s}\right|}{2\eta}-s\frac{\gamma_{s}}{2\eta}\right)$, we can apply the Taylor expansion in the expression (\ref{2.15}) up to the first order terms, and find
\begin{eqnarray}
\mathcal{E}_{n,\,l}\approx m+2\frac{dE_{0}\omega\eta}{m}\left[n+\frac{\left|\gamma_{s}\right|}{2\eta}+s\frac{\gamma_{s}}{2\eta}+1\right]+s\,d\,E_{0}-\omega\left[l+\frac{1}{2}\right],
\label{2.20}
\end{eqnarray}
where the first term of the right-hand-side of Eq. (\ref{2.20}) corresponds to the rest mass of the neutral particle. The other terms correspond to the nonrelativistic analogue of the Landau-He-McKellar-Wilkens quantization obtained in \cite{b} under the influence of the topology of a disclination. We can see that the cyclotron frequency is given by $\omega_{c}=2\frac{d\,E_{0}\,\omega\,\eta}{m}$, that is, the cyclotron frequency depends on the intensity of the electric field, the angular velocity of the noninertial frame and the parameter $\eta$, associated with the deficit of angle \cite{b}. This dependence of the cyclotron frequency on the parameter $\eta$ comes from the noninertial effects. We also recover the term due to Page-Werner \emph{et al}, given by the coupling between the angular velocity $\omega$ and the quantum number $l$ \cite{r1,r2,r4}.

\section{Dirac spinors and the Gordon decomposition}

We now turn to the Dirac spinors corresponding to the positive-energy solutions of the Dirac equation (\ref{2.6}). The appropriate solutions can be obtained by solving the system of coupled equation given in (\ref{2.8}) and (\ref{2.9}). We have already obtained the solutions for the two-spinor $\xi$. Their radial eigenfunctions are given by the expression
\begin{eqnarray}
\xi_{s}\left(\rho\right)=\left(d\,E_{0}\,\omega\,\eta\right)^{\frac{\left|\zeta_{s}\right|}{2\eta}}\,e^{\frac{-d\,E_{0}\,\omega\,\eta\,\rho^{2}}{2}}\,\rho^{\frac{\left|\zeta_{s}\right|}{\eta}}\,F\left[-n,\frac{\left|\zeta_{s}\right|}{\eta}+1,d\,E_{0}\,\omega\,\eta\,\rho^{2}\right]\qquad(s=\pm).
\label{eq:2.201}
\end{eqnarray} 

We now substitute these solutions into (\ref{2.9}) to obtain the solutions for the two-spinor $\chi$. The positive-energy solution of the Dirac equation (\ref{2.6}) running parallel to the $z$ axis is
\begin{eqnarray}
\psi_{+}&=&g_{+}\,F\left[-n,\frac{\left|\zeta_{+}\right|}{\eta}+1,d\,E_{0}\,\omega\,\eta\,\rho^{2}\right]\left(
\begin{array}{c}
1 \\
0\\
0\\
-\frac{i}{\left[\mathcal{E}+m+\omega\left(l+\frac{1}{2}\right)-d\,E_{0}\right]}\left(\frac{\left|\zeta_{+}\right|}{\eta\rho}-\frac{\zeta_{+}}{\eta\rho}-2d\omega E_{0}\eta\rho\right)\\	
\end{array}\right)\nonumber\\
[-3mm]\label{2.17}\\[-3mm]
&+&\frac{i\,g_{+}}{\left[\mathcal{E}+m+\omega\left(l+\frac{1}{2}\right)-d\,E_{0}\right]}\,F\left[-n+1,\frac{\left|\zeta_{+}\right|}{\eta}+2,d\,E_{0}\,\omega\eta\,\rho^{2}\right]\,\left(
\begin{array}{c}
0\\
0\\
0\\
\frac{2n\,d\,E_{0}\,\omega\,\eta\,\rho}{\left(\frac{\left|\zeta_{+}\right|}{\eta}+1\right)}\\	
\end{array}\right),\nonumber
\end{eqnarray}
while the positive-energy solution antiparallel to the $z$ axis is
\begin{eqnarray}
\psi_{-}&=&g_{-}\,F\left[-n,\frac{\left|\zeta_{-}\right|}{\eta}+1,d\,E_{0}\,\omega\,\eta\,\rho^{2}\right]\left(
\begin{array}{c}
0 \\
1\\
-\frac{i}{\left[\mathcal{E}+m+\omega\left(l+\frac{1}{2}\right)+d\,E_{0}\right]}\left(\frac{\left|\zeta_{-}\right|}{\eta\rho}+\frac{\zeta_{-}}{\eta\rho}-2d\omega\eta E_{0}\rho\right)\\	
0\\
\end{array}\right)\nonumber\\
[-3mm]\label{2.18}\\[-3mm]
&+&\frac{i\,g_{-}}{\left[\mathcal{E}+m+\omega\left(l+\frac{1}{2}\right)+d\,E_{0}\right]}\,F\left[-n+1,\frac{\left|\zeta_{-}\right|}{\eta}+2,d\,E_{0}\,\omega\,\eta\,\rho^{2}\right]\,\left(
\begin{array}{c}
0\\
0\\
\frac{2n\,d\,E_{0}\,\omega\,\eta\,\rho}{\left(\frac{\left|\gamma_{-}\right|}{\eta}+1\right)}\\	
0\\
\end{array}\right).\nonumber
\end{eqnarray}
The prefactors $g_{\pm}$ in Eqs.~(\ref{2.17})~and (\ref{2.18}) are defined by the equality
\begin{eqnarray}
g_{\pm}=C\,e^{-i\mathcal{E}t}\,e^{i\left(l+\frac{1}{2}\right)\varphi}\,e^{ikz}\,\left(d\,E_{0}\,\omega\,\eta\right)^{\frac{\left|\zeta_{\pm}\right|}{2\eta}}\,e^{\frac{-d\,E_{0}\,\omega\,\eta\,\rho^{2}}{2}}\,\rho^{\frac{\left|\zeta_{\pm}\right|}{\eta}},
\label{2.19}
\end{eqnarray}
where $C$ is a constant. 

The spinors~(\ref{2.17})~and (\ref{2.18}) are positive-energy solutions of the Dirac equation~(\ref{2.6}). The same procedure leads to the negative-energy solutions. All components of the spinors depend on the parameter $\eta$ from the cosmic string spacetime. In the limit $\eta\rightarrow1$, the Dirac spinors for positive-energy solutions in the Minkowski spacetime obtained in \cite{b3} are recovered.

Finally, let us discuss the Gordon decomposition \cite{gordon,greiner}. The Gordon decomposition \cite{gordon,greiner} is applied to the Dirac spinor field $\psi$, and consists in splitting the 4-vector $J^{\mu}=\bar{\psi}\gamma^{\mu}\psi$ up into the convection current density $J^{\mu}_{\mathrm{conv}}=-\frac{i}{2m}g^{\mu\nu}\,\left[\bar{\psi}\partial_{\nu}\psi-\left(\partial_{\nu}\bar{\psi}\right)\psi\right]$ and the spin-current density $J^{\mu}_{\mathrm{spin}}=\frac{1}{2m}\partial_{\nu}\left[\bar{\psi}\Sigma^{\mu\nu}\psi\right]$. In a curved spacetime background, the components of $J^{\mu}$ are given by \cite{gordon2}
\begin{eqnarray}
J^{\mu}&=&-\frac{i}{2m}\,g^{\mu\nu}\,\left[\bar{\psi}\partial_{\nu}\psi-\left(\partial_{\nu}\bar{\psi}\right)\psi\right]-\frac{1}{2m}\,\partial_{\nu}\left[\bar{\psi}\Sigma^{\mu\nu}\psi\right]+\frac{i}{2m}\,\bar{\psi}\left[\gamma^{\nu}\Gamma_{\nu},\gamma^{\mu}\right]\psi\nonumber\\
&-&\frac{i}{4m}\bar{\psi}\left[\left(\partial_{\nu}\gamma^{\nu}\right),\gamma^{\mu}\right]\psi-\frac{i}{4m}\left[\gamma^{\nu},\left(\partial_{\nu}\gamma^{\mu}\right)\right]\psi.
\label{2.20}
\end{eqnarray}

Simple calculations then lead to the following expressions for the components of $J^{\mu}$:
\begin{eqnarray}
J^{t}&=&J^{t}_{\mathrm{conv}}-\vec{\nabla}\cdot\vec{P}\nonumber\\
J^{\rho}&=&J^{\rho}_{\mathrm{conv}}+\left(\frac{\partial}{\partial t}-\omega\frac{\partial}{\partial\varphi}\right)P^{\rho}-\left(\vec{\nabla}\times\vec{M}\right)^{\rho}\nonumber\\
[-2mm]\label{2.21}\\[-2mm]
J^{\varphi}&=&J^{\varphi}_{\mathrm{conv}}+\left(\frac{\partial}{\partial t}-\omega\frac{\partial}{\partial\varphi}\right)P^{\varphi}-\left(\vec{\nabla}\times\vec{M}\right)^{\varphi}+\frac{M^{z}}{\eta\rho^{2}}+\omega\,\vec{\nabla}\cdot\vec{P}\nonumber\\
J^{z}&=&J^{z}_{\mathrm{conv}}+\left(\frac{\partial}{\partial t}-\omega\frac{\partial}{\partial\varphi}\right)P^{z}-\left(\vec{\nabla}\times\vec{M}\right)^{z}\nonumber,
\end{eqnarray}
where we have defined in (\ref{2.21}) the magnetization current density $M^{\mu}=e^{\mu}_{\,\,\,a}\left(x\right)M^{a}$, whose components of $M^{a}$ are 
\begin{eqnarray}
M^{1}=\frac{1}{2m}\bar{\psi}\Sigma^{1}\psi;\,\,\,\,M^{2}=\frac{1}{2m}\bar{\psi}\Sigma^{2}\psi;\,\,\,\,M^{3}=\frac{1}{2m}\bar{\psi}\Sigma^{3}\psi,
\label{2.23}
\end{eqnarray}
and the polarization density $P^{\mu}=e^{\mu}_{\,\,\,a}\left(x\right)P^{a}$, whose components of $P^{a}$ are 
\begin{eqnarray}
P^{1}=\frac{i}{2m}\bar{\psi}\gamma^{0}\gamma^{1}\psi;\,\,\,\,P^{2}=\frac{i}{2m}\bar{\psi}\gamma^{0}\gamma^{2}\psi;\,\,\,\,P^{3}=\frac{i}{2m}\bar{\psi}\gamma^{0}\gamma^{3}\psi.
\label{2.22}
\end{eqnarray}

The topology of the cosmic string and the noninertial effects on the Fermi-Walker reference frame contribute to the spatial components of the current $J^{\mu}$. In the limit $\eta\rightarrow1$, we recover the results in Ref.~\onlinecite{b3} for the components of $J^{\mu}$ in the Fermi-Walker reference frame in the Minkowski spacetime. In the limit $\omega\rightarrow0$ with $0<\eta<1$, Eq.~(\ref{2.21}) determines each $J^{\mu}$ component in an inertial frame in the cosmic string spacetime. In the limits $\eta\rightarrow1$ and $\omega\rightarrow0$, we recover the expressions for the $J^{\mu}$ in an inertial frame in the Minkowski spacetime.

\section{Conclusions}

We have discussed the behaviour of the external fields interacting with a Dirac neutral particle with a permanent electric dipole moment in a noninertial frame in the cosmic string spacetime. We have shown that the noninertial effects of the Fermi-Walker reference frame induces a radial magnetic field by considering initially a uniform electric field along the $z$ axis in the rest frame of the observers. Among the interesting aspects of the induced radial magnetic field is its dependence on the parameter $\eta$, directly related to the deficit of angle of the defect. In the field configuration induced by the noninertial effects, the electrostatic conditions are satisfied, no torque acts on the electric dipole moment, and a uniform magnetic field is generated in the direction perpendicular to the plane on which the Dirac particle moves. Under these conditions, we have shown that bound states solutions of the Dirac equation can only be achieved by restricting the intensity of the electric field to $E_{0}\ll\frac{\omega\eta}{d}$. For other values of the intensity of the electric field, the amplitude of probability cannot be considered sufficiently small in order to have the wave function being normalized inside the physical region of the cosmic string spacetime. The relativistic bound states solutions correspond to the analogue of the Landau quantization for a Dirac neutral particle with a permanent electric dipole moment in the cosmic string spacetime which is called the relativistic Landau-He-McKellar-Wilkens quantization.

We have also seen that influence of the topology of the defect on the relativistic energy levels yields the breaking of the degeneracy of the relativistic Landau levels. Nonetheless, in the limit $\eta\rightarrow1$, the relativistic Landau-He-McKellar-Wilkens induced by noninertial effects in the Minkowski spacetime is recovered. By contrast with a previous work \cite{bf5}, we have shown that the relativistic Landau-He-McKellar-Wilkens quantization induced by noninertial effects can be achieved without assuming the existence of a density of magnetic charges.

Finally, we have calculated the components of the Dirac spinor that are parallel and anti-parallel to the spacetime $z$-axis for positive-energy solutions and obtained the Gordon decomposition of the Dirac probability current. The results show that the noninertial effects of the Fermi-Walker reference frame and the topology of the cosmic string spacetime influence the Dirac spinors and the Gordon decomposition.

The author would like to thank CNPq (Conselho Nacional de Desenvolvimento Cient\'ifico e Tecnol\'ogico - Brazil) for financial support.

\end{document}